\documentclass[%
 reprint,
preprintnumbers,
nofootinbib,
amsmath,amssymb,aps,superscriptaddress
]{revtex4-2}
\counterwithout{equation}{section}
\usepackage{amsmath,float,graphicx,placeins,amssymb,multirow,pgfplots,pgfplotstable}
\usepackage{empheq}
\usepackage{booktabs}
\usepackage{tikz}
\usepackage{amsmath}
\usepackage[most]{tcolorbox} % robust boxing
\usepackage{amsmath,amssymb,amsfonts,bm}
\usepackage{mathtools}
\usepackage{physics}
\usepackage{microtype}
\usepackage{xcolor}
\usepackage{comment}
\usepackage{enumerate,slashed,cancel,comment,color,bbm,graphicx,rotating,placeins,mathtools,multirow,hhline}
\usepackage[normalem]{ulem}
\usepackage{array}
\usepackage{hyperref}
\usepackage{color}
\usepackage{booktabs}
 \hypersetup
 {
   colorlinks,
   linkcolor={blue!80!black},
   citecolor={blue!70!black},
   urlcolor={blue!70!black}
 }

\usepackage{hyperref}
\usepackage{tabularx}

\newcommand{\be}{\begin{equation}}
\newcommand{\ee}{\end{equation}}
\newcommand{\bea}{\begin{eqnarray}}
\newcommand{\eea}{\end{eqnarray}}
\newcommand{\ba}{\begin{aligned}}
\newcommand{\ea}{\end{aligned}}

\pgfplotsset{compat=1.18} 
\begin{document}
\preprint{KCL-PH-TH/2025-47}
\title{Effect of the Memory Burden on Primordial Black Hole Hot Spots}
% \nate{[not 100 percent on the title, it seems a little cumbersome, though it's succinctness is very good. perhaps "Effect of the memory burden on primordial black hole hot spots" or "primordial black hole hotspots and the effect of the memory burden" what do you think?]}

\begin{abstract} 
When primordial black holes (PBHs) evaporate, they deposit energy in the surrounding plasma, leading to temperature gradients, or {\it hot spots}, that evolve during the evaporation process. Motivated by recent studies suggesting that a memory burden may slow down PBH evaporation, we explore how a suppression of the evaporation rate affects the morphology of such hot spots. We include such a suppression in the form of transfer functions and derive general formulas for the hot-spot core temperature and radius. Applying our results to illustrative scenarios, we find that in the vanilla memory burden scenario in which the evaporation rate and Hawking temperature are exactly constant, the hot-spot temperature is substantially lowered. Nonetheless, we show that alternative scenarios may lead to sizeable hot spots with morphologies that differ significantly from the semi-classical case.
\end{abstract}

\author{Nathaniel Levy}
\email[Email address: ]{nathaniel.levy@stx.ox.ac.uk}

\affiliation{Rudolf Peierls Centre for Theoretical Physics, University of Oxford,\\ Oxford OX1 3PU, United Kingdom}
\affiliation{Theoretical Particle Physics and Cosmology, King’s College London,\\ Strand, London WC2R 2LS, United Kingdom}

\author{Lucien Heurtier}
\email[Email address: ]{lucien.heurtier@kcl.ac.uk}

\affiliation{Theoretical Particle Physics and Cosmology, King’s College London,\\ Strand, London WC2R 2LS, United Kingdom}

\maketitle

\section{Introduction}
Primordial Black Holes (PBHs) that formed in the early universe, besides providing an alternative candidate to the dark matter of our universe~\cite{Carr:2020gox, Green:2020jor}, constitute some of the very few remnants that could have survived across cosmological time and left imprints in cosmological data that might be measurable today. PBHs form from the collapse of primordial curvature over-densities \cite{Carr:1975qj}, which can originate from various physical mecanisms~\cite{Carr:1993aq,Ivanov:1994pa,Garriga:2015fdk,Deng:2016vzb,Deng:2017uwc,Kusenko:2020pcg,Heurtier:2022rhf, Ai:2024cka, Maeda:1981gw,Sato:1981gv,Kodama:1981gu,Kodama:1982sf,Hsu:1990fg,Liu:2021svg,Kawana:2022olo,Gouttenoire:2023naa,Hawking:1982ga,Moss:1994iq,Moss:1994pi, Khlopov:1998nm, Crawford:1982yz,Gross:2021qgx,Baker:2021sno,Kawana:2021tde, Dolgov:1992pu,Dolgov:2008wu,Kitajima:2020kig,Kasai:2022vhq,Martin:2019nuw,Martin:2020fgl,Rubin:2000dq, Vachaspati:2017hjw,Ferrer:2018uiu,Gelmini:2022nim,Gelmini:2023ngs,Hawking:1987bn,Polnarev:1988dh,Fort:1993zb,
Garriga:1993gj,Caldwell:1995fu,MacGibbon:1997pu,Jenkins:2020ctp,Blanco-Pillado:2021klh}.  As such, they may also constitute unique probes of very-high energy particle physics phenomena. Nevertheless, PBHs with masses less than $10^{15}$ g evaporate on time scales smaller than the age of the Universe, due to Hawking radiation, which makes hunting for their existence in cosmology very challenging. 

So far, studies searching for the signatures of evaporating PBHs in cosmological data  (see, e.g., \cite{Auffinger:2022khh} for a review) assumed that their evaporation is a homogeneous process. However, it was suggested in Refs.~\cite{He:2022wwy,He:2024wvt,Das:2021wei}, that the dynamics of energy deposition onto the plasma around PBHs when they evaporate can lead to the formation of {\em hot spots} with large local temperature gradients. Since then, such gradients were revealed to be extremely important when assessing the effect of PBH evaporation on particle physics phenomenology~\cite{Gunn:2024xaq, Altomonte:2025hpt, Hamaide:2023ayu}.

One crucial ingredient in obtaining the temperature profile time evolution around PBHs is the PBH evaporation rate, that was derived by S. Hawking in the semi-classical (SC) approximation~\cite{Hawking:1974rv,Hawking:1974sw}. In Refs.~\cite{He:2022wwy,He:2024wvt,Das:2021wei}, this evaporation rate was used to obtain the time-evolution of PBH hot spots during their evaporation. Recently, it was proposed that this evaporation rate is affected by the {\em memory burden}(MB) effect, a property of high-entropy storage quantum systems that may be featured by black holes, affecting their evaporation dynamics~\cite{Dvali:2018xpy,Dvali:2018ytn,Dvali:2020wft}. Remarkably, this effect---which can be thought of as the backreaction of the metric evolution on the particle production rate---can significantly prolong the life of a black hole, resulting in a markedly different PBH phenomenology and corresponding strategies to search for their existence in the cosmos
% } tends to significantly increase the lifetime of a black hole, leading to a completely different phenomenology of PBHs and fairly different ways to search for their evaporation in the cosmos
~\cite{Montefalcone:2025akm,Kitabayashi:2025iaq,Dvali:2025ktz,Chianese:2024rsn,Kohri:2024qpd,Bhaumik:2024qzd,Alexandre:2024nuo,Tseng:2025fjf,Chaudhuri:2025asm,Dondarini:2025ktz,Tan:2025vxp,Chaudhuri:2025rcs,Chianese:2025wrk,Liu:2025vpz,Calabrese:2025sfh,Bandyopadhyay:2025ast,Athron:2024fcj,Borah:2024bcr,Barman:2024iht,Barman:2024ufm,Gross:2025hia,Gross:2024wkl,Haque:2024eyh}.

In this work, we investigate how this memory burden effect---or, more generally, a modification of the SC rate derived by Hawking---may lead to a different morphology and dynamics for PBH hot spots. The paper is organised as follows: In Sec.~\ref{sec:evaporation} we introduce the definition of the semi-classical evaporation and its memory-burden modification. In Sec.~\ref{sec:hotspots}, we review the results of Refs.~\cite{He:2022wwy,He:2024wvt,Das:2021wei} in the SC case. Finally, in Sec.~\ref{sec:MBhotspots}, we generalise these results in the presence of the memory burden, and study how the PBH hot-spot morphology is affected by this modified PBH dynamics, before concluding in Sec.~\ref{sec:conclusion}.
\section{PBH Evaporation}\label{sec:evaporation}
We start by reviewing standard results regarding the evaporation of primordial black holes in the early universe. We then report how the memory burden effect was proposed to be accounted for in the literature, so we set the scene and basic ingredients that we will then use in the next section when studying the dynamics of PBH hot spots.
\subsection{Semi-Classical Approximation}
Assuming that Hawking radiation takes place on top of a classical background that is uniquely determined by the total mass of the black hole---the SC approach---Hawking~\cite{Hawking:1975vcx} showed that all available degrees of freedom evaporated by a PBH of mass $M$ follow a blackbody spectrum distribution with temperature
\begin{align}\label{eq:HawkTemp}
T_H = \frac{1}{8\pi\, G M}\,\sim 10^3~{\rm TeV} \left(\frac{10^7 g }{M}\right)\,,
\end{align}
where $G$ denotes the Newton constant. The differential emission rate per unit energy $E$, time $t$, of a particle species $i$ with mass $m_i$ is given by
\begin{equation}
 \frac{d^{2}N_i}{dt\, dE} = \frac{g_i}{2\pi}\frac{\vartheta(M,E)}{e^{E/T_{H}}-(-1)^{2s_i}}\,,\label{eq:emission_rate}
\end{equation}
%%%
where $s_i$ denotes the spin of the particle species $i$ and $g_i$ its number of internal degrees of freedom, and the absorption cross section $\vartheta(M,E)$ represents the likelihood that a particle reaches spatial infinity, accounting for the curvature-induced potential barrier acting on particles around the black hole. The PBH mass loss rate in the semi-classical (SC) case can then be estimated as~\cite{Hawking:1975vcx, MacGibbon:1990zk, MacGibbon:1991tj}
%%%
\begin{align} \label{eq:MEq}
 \left.\frac{dM}{dt}\right|_{\rm SC} = -\sum_i \int_{m_i}^{\infty} \frac{d^{2}N_i}{dt\, dE}\, E\, dE = - \varepsilon(M)\, \frac{m_p^4}{M^2}\,,
\end{align}
%%%
where $m_p\approx1.22\times 10^{22}\mathrm{GeV}$ denotes the Planck mass and $\varepsilon(M)$ is the evaporation function that accounts for the degrees of freedom that can be produced at a given time, i.e., for an instantaneous mass. Although, in principle, this evaporation function is a function of the mass whose variation needs to be tracked numerically (see for instance Ref.~\cite{Cheek:2021odj}), at high energy this evaporation function is constant $\varepsilon\approx 4.2\times 10^{-3}$, which we will assume throughout this work. 

\subsection{Memory Burden Effect}
When accounting for the memory burden \cite{Dvali:2020wft, Dvali:2024hsb}, one usually considers that the evaporation rate follows Eq.~\eqref{eq:MEq} from the time of PBH formation, when $M=M_{\rm ini}$, to the point when the memory burden kicks in, usually assumed to be an order one fraction of the initial mass at $M=qM_{\rm ini}$ where $q\lesssim 1$. After that, the effect is typically parametrised as
\be
 \label{eq:MEqMB}
 \left.\frac{dM}{dt}\right|_{\rm MB} = \frac{1}{\tilde S^k} \left.\frac{dM}{dt}\right|_{\rm SC}\,, 
\ee
where $k\geqslant 0$ (with $k=0$ reproducing the SC result) and the suppression factor is directly inferred from the Hawking-Bekenstein entropy of the PBH,
\be
\tilde S(M)\equiv \frac{S}{k_{\mathrm{B}}}
  = \frac{4 \pi G M^2}{\hbar k_{\mathrm{B}}}
  \simeq 2.6 \times 10^{30}
  \left( \frac{M}{10^{10}\,\mathrm{g}} \right)^{2},
\label{eq:entropy_dimless}
\ee
where $k_{\mathrm{B}}$ is the Boltzmann constant. 

Although it was argued in Ref.~\cite{Dvali:2025ktz} that in Eq.~\eqref{eq:MEqMB} the rate should be understood as being constant across the rest of the evaporation, its interpretation in the literature has been diverse, with some considering this expression as holding true at any time, such that $S(M)$ and $dM/dt|_{\rm SC}$ both vary over time with $M(t)$ ~\cite{Haque_2024,ettengruber2025microblackholedark,Thoss_Burkert_Kohri_2024b,Balaji2024ProbingMH} and some considering the rate to be constant~\cite{Montefalcone:2025akm,Chaudhuri:2025asm,Thoss_Burkert_Kohri_2024b,Balaji2024ProbingMH}. 
% \lucien{add citations}\nate{I have added two for each, is this enough?}

To account for this diversity of interpretations, and the fact that the exact evolution of the Hawking temperature and the evaporation rate might deviate from these extreme cases when using a more sophisticated approach, we encapsulate the MB effect by considering that these two crucial quantities evolve as
\be\label{eq:MBparam}
\left.\frac{dM}{dt}\right|_{\rm MB} \equiv \eta_M\times \left.\frac{dM}{dt}\right|_{\rm SC}\ \text{and}\quad T_H\equiv \eta_T \times T_{H,\mathrm{SC}}\,,
\ee
where $T_{H,\mathrm{SC}}$ refers to the semi-classical Hawking temperature formula used in Eq.~\eqref{eq:HawkTemp}.

In the true MB scenario where both the evaporation rate and Hawking temperature freeze at $M=qM_{\rm ini}$, these functions correspond to 
\bea\label{eq:etaM}
\eta_M=\left\{\begin{matrix}
1\,,\quad\quad\quad\quad\quad
qM_{\rm ini}<M\leqslant M_{\rm ini}\,,\\
~\\
\displaystyle\frac{\displaystyle 1}{\displaystyle\tilde S(qM_{\rm ini})^k}\left(\displaystyle\frac{M}{qM_{\rm ini}}\right)^2\,,\quad M\leqslant qM_{\rm ini}\,,
\end{matrix}\right.
\eea
and
\bea\label{eq:etaT}
\eta_T=\left\{\begin{matrix}
1\,,\quad\quad qM_{\rm ini}<M\leqslant M_{\rm ini}\,,\\
~\\
\displaystyle\frac{M}{qM_{\rm ini}}\,,\quad\quad\quad\quad M\leqslant qM_{\rm ini}\,.
\end{matrix}\right.
\eea
In scenarios incorporating a smooth transition between the SC and MB rate, but maintaining a sharp transition for the Hawking temperature, such as in Refs.~\cite{Montefalcone:2025akm, Dvali:2025ktz}, $\eta_T$ would remain the same, but $\eta_M$ would typically be a smoother function of the PBH mass, or of time, depending on how it is parametrised.

% in what follows, we will consider three gradual variations of these two opposite cases:\lucien{add citations}
% \begin{itemize}
%     \item {\bf Case 1:} The whole rate is evaluated at $M=qM_{\rm ini}$ throughout the rest of the evaporation;
%     \item {\bf Case 2:} Only the memory-burden suppression factor $\tilde S(qM_{\rm ini})$ is taken to be constant, but the SC rate is take to vary in the usual way with $M(t)$;
%     \item {\bf Case 3:} The whole rate---with both the SC rate and the suppression factor $\tilde S(M)$---is assumed to vary in a self-similar manner when $M(t)$ evolves with time.
% \end{itemize}

\section{PBH Hot Spots}\label{sec:hotspots}

During their evaporation, PBHs inject energy in the surrounding plasma, creating temperature gradients and supporting the evolution of hot spots with a characteristic temperature profile around them. The existence of such hot spots was first suggested in Ref.~\cite{Das:2021wei} and refined--accounting for the Landau-Pomeranchuk-Migdal (LPM) effect in the calculation of the Hawking radiation energy deposition rate in Ref.~\cite{He:2022wwy}---before it was supported by a Boltzmann-improved numerical treatment~\cite{He:2024wvt}. In these references, the plasma is assumed to be in local thermal equilibrium, such that it can be uniquely described at every radius $r$ by a scalar function $T(r)$. The rate of energy deposition of high-energy Hawking radiation particles with momentum $k\approx T_H$ onto a much lower-energy plasma of temperature $T\ll T_H$ is given by the Landau–Pomeranchuk–Migdal (LPM) rate~\cite{He:2022wwy}
\be 
\Gamma_{\rm dep}(T,T_H) \sim \alpha^2T\sqrt{\frac{T}{T_H}}\,.
\ee
The diffusion rate that encodes the radial cooling of the profile is of the Bethe-Heitler form
\be 
\Gamma_{\rm diff}(T)\sim \alpha^2 T\,,
\ee
where in these two expressions, the coupling constant $\alpha \sim 0.1$ is assumed to be universal, which is a good approximation at high energy when the evaporation is dominated by hard gluons~\cite{He:2022wwy, Mukaida:2022bbo}.

\subsection{Hot-Spot Morphology}

To obtain the temperature gradient profile within the hot spot, one first needs to determine, at a given time, under which radius the thermal plasma is able to thermalise faster than it is being heated by PBH evaporation. To do so, one can proceed in three steps: First, consider that Hawking radiation deposits energy onto a plasma of temperature $T$ at the distance $\Gamma_{\rm dep}(T,T_H)^{-1}$ away from the BH. Then, estimate the time $t_{\rm diff}$ it takes within a plasma of temperature $T$ for a plasma particle interacting with a diffusion rate $\Gamma_{\rm diff}(T)$ to propagate over a similar distance. Using that  $r_{\rm diff}\sim \sqrt{t_{\rm diff}/\Gamma_{\rm diff}}$, this gives~\cite{He:2022wwy,He:2024wvt,Das:2021wei}
\be
r_{\rm diff}=\Gamma_{\rm dep}^{-1}\ \Rightarrow\ t_{\rm diff}=\alpha^{-2}T^{-2}\frac{m_p^2}{8\pi M}\,.
\ee 
Finally, demanding that, within that time, the energy contained in a sphere of radius $\Gamma_{\rm dep}(T,T_H)^{-1}$ equals the energy deposited by Hawking radiation provides an expression for the hot-spot core temperature~\cite{He:2022wwy,He:2024wvt,Das:2021wei}
\be
\frac{\pi^2}{30}g_{\star}(T_c)T_c^4\times \frac{4\pi}{3}\Gamma_{\rm dep}(T_c,T_H)^{-3}=-t_{\rm diff}\frac{dM}{dt}
\ee
Using the expression of the Hawking temperature and evaporation rate in the presence of MB in Eq.~\eqref{eq:MBparam}, we obtain the core temperature, as well as the corresponding radius of the core $r_c$,
\noindent\refstepcounter{equation}\label{eq:Tcrc}
\begin{tcolorbox}[enhanced, colback=white, colframe=black, boxrule=0.5pt,
                  left=3pt, right=3pt, top=-6pt, bottom=2pt]
  \[
  \begin{aligned}
      T_c &= \left(\frac{\eta_M^2}{\eta_T}\right)^{1/3}T_{c,\mathrm{SC}}\,,\\
      r_c &=\left(\frac{\eta_T}{\eta_M}\right)\ r_{c,\mathrm{SC}}\,,
  \end{aligned}
  \tag*{\((\theequation)\)} % <- prints the number inside the box
  \]
\end{tcolorbox}
where the semi-classical results were derived in Ref.~\cite{He:2022wwy}
\bea
T_{c,\mathrm{SC}}&=&24\left(\frac{150}{\pi ^2}\right)^{1/3}\alpha ^{8/3} \left(\frac{\varepsilon}{g_\star}\right)^{2/3}T_{H,\mathrm{SC}}\\
&\approx& 1.5\times 10^{-4}\left(\frac{\alpha}{0.1}\right) ^{8/3} \times\nonumber\\
&&\left(\frac{\varepsilon}{4.2\times 10^{-3}}\right)^{2/3}\!\!\left(\frac{g_\star}{106.75}\right)^{-2/3}\!T_{H,\mathrm{SC}}\,.
\eea
and
\bea
&&r_{c,\mathrm{SC}}=\frac{\pi^2 g_\star}{360 \alpha ^6 \varepsilon}r_{H}\,,\nonumber\\
&\approx & 7.0\times 10^8 \left(\frac{\alpha}{0.1}\right)^{\!-6}\! \left(\frac{\varepsilon}{4.2\times 10^{-3}}\right)^{\!-1}\!\!\left(\frac{g_\star}{106.75}\right)r_{H}\,,\nonumber\\
\eea
in which \mbox{$r_H=2M/m_p^2$} denotes the classical Schwarzchild radius of the BH and $g_\star$ needs, in principle, to be evaluated at the true hot-spot core temperature $T_c$. In practise, like we did for the evaporation function $\varepsilon$, we will assume this number of relativistic degrees of freedom to be constant throughout this work.

Outside this homogeneous core, a local thermal equilibrium persists, but diffusion is at play, with a temperature evolving as $\propto r^{-1/3}$~\cite{He:2022wwy}, which remains the case in the presence of memory burden. 
During the evaporation time scale\footnote{Note that this time scale differs with an $\mathcal O(1)$ prefactor from the total evaporation time of the PBH.}
\be
\tau_{\rm ev}\equiv\left(\frac{1}{M}\frac{dM}{dt}\right)^{-1}\,,
\ee
such diffusion can only propagate until the maximum radius
\be
r_{\rm dec}\equiv\sqrt{\frac{\tau_{\rm ev}}{T(r_{\rm dec})}}\,,\quad T(r_{\rm dec})=T_c(r_c/r_{\rm dec})^{-1/3}\,,
\ee
giving
\noindent\refstepcounter{equation}\label{eq:rdec}
\begin{tcolorbox}[enhanced, colback=white, colframe=black, boxrule=0.5pt,
                  left=3pt, right=3pt, top=-8pt, bottom=2pt]
  \[
  \begin{aligned}
      r_{\rm dec} &= \eta_M^{-4/5}r_{\mathrm{dec,SC}}\,,\\
  \end{aligned}
  \tag*{\((\theequation)\)} % <- prints the number inside the box
  \]
\end{tcolorbox}
where~\cite{He:2022wwy}
\bea
&& r_{\mathrm{dec, SC}}=\frac{\pi ^{3/5} g_\star^{1/5}}{720^{1/5} \alpha ^{8/5} \varepsilon^{4/5}}\left(\frac{M}{m_p}\right)^{6/5}r_H\,,\\
&=&1.1\times 10^{20} \left(\frac{\alpha}{0.1}\right)^{\!-8/5}\! \left(\frac{\varepsilon}{4.2\times 10^{-3}}\right)^{\!-4/5}\!\!\nonumber\\
&&\left(\frac{g_\star}{106.75}\right)^{1/5}\left(\frac{M}{10^9\mathrm{g}}\right)^{6/5}r_{H}\,,
\eea
For $r_{\rm dec}(M)<r<r_{\rm dec}(M_{\rm ini})$, though diffusion still contributes, it is no longer efficient due to the increasingly short timescales. Consequently, in Ref.~\cite{He:2022wwy}, it was shown that the temperature decreases like $\propto r^{-7/11}$, such that the overall temperature profile is
\be
T=\left\{\begin{matrix}
T_c\,,\quad\quad\quad\quad\quad\quad\quad r_H<r\leqslant r_c \,,\\
~\\
T_c\left(\displaystyle\frac{r}{r_c}\right)^{-1/3}\,,\quad\ \quad\quad r_c<r\leqslant r_{\rm dec}\,,\\
~\\
T_c\!\left(\displaystyle\frac{r_{\rm dec}}{r_c}\right)^{\!\!-1/3}\!\!\left(\displaystyle\frac{r}{r_{\rm dec}}\right)^{\!\!-7/11}\!\!\!\!\!\!\!\!\!, \quad r_{\rm dec}<r\leqslant r_{\rm dec}(qM_{\rm ini})\,.
\end{matrix}\right.
\ee
Note that these equations remain valid unless $r_c\geqslant r_{\rm dec}$, after which diffusion is no longer able to establish a local thermal equilibrium on scales smaller than the core. This happens when the mass decreases below a critical mass, 
% \vspace{-10pt}
\noindent\refstepcounter{equation}\label{eq:Mstar}
\begin{tcolorbox}[enhanced, colback=white, colframe=black, boxrule=0.5pt,
                  left=3pt, right=3pt, top=-8pt, bottom=2pt]
  \[
  \begin{aligned}
      M_\star = \eta_M^{-1/6}\eta_T^{5/6}M_{\star,\mathrm{SC}}
\,,
  \end{aligned}
  \tag*{\((\theequation)\)} % <- prints the number inside the box
  \]
\end{tcolorbox}\noindent
where $\eta_M$ and $\eta_T$ need to be evaluated at $M_\star$ and the semi-classical value is\footnote{Note that there is a factor of $3^{5/12}$ of difference between this result and the result obtained in ~\cite{He:2022wwy}, as we have used $\tau_{\rm ev}$ in the evaluation of $r_{\rm dec}$ as opposed to the integrated PBH lifetime.}~\cite{He:2022wwy} 
\bea
&&M_{\star,\mathrm{SC}}= \frac{\pi ^{7/6}}{(64800 \sqrt{2})^{1/3}}\frac{g_\star^{2/3}m_p}{\alpha^{11/3}\varepsilon^{1/6}}\,,\nonumber\\
&\approx& 0.58g \left(\frac{\alpha}{0.1}\right)^{\!-11/3}\! \left(\frac{\varepsilon}{4.2\times 10^{-3}}\right)^{\!-1/6}\!\!\left(\frac{g_\star}{106.75}\right)^{2/3}\,.\nonumber\\
\eea
Once the PBH mass drops below this critical mass, it was argued in Ref.~\cite{He:2022wwy} that the hot spot does not vary by much since the remaining energy evaporated by the black hole is subdominant in comparison to total energy of the hot-spot core. This energy does not efficiently diffuse within the hot spot, and we will assume this to be true in this study.
% Once the mass of the PBH decreases below that mass, it was argued in Ref.~\cite{He:2022wwy} that the hot spot does not vary much, as the remaining energy evaporated by the black hole is subdominant compared to the hot spot core total energy and this energy doesn't efficiently diffuse within the hot spot, which we will assume true in this study.

At $M=M_\star$, the PBH has thus reached its maximum core temperature, equal to 
\noindent\refstepcounter{equation}\label{eq:Tmax}
\begin{tcolorbox}[enhanced, colback=white, colframe=black, boxrule=0.5pt,
                  left=3pt, right=3pt, top=-8pt, bottom=2pt]
  \[
  \begin{aligned}
      T_{\rm max}&=&\eta_M^{5/6}\eta_T^{-7/6}T_{\rm max, SC}
\,,
  \end{aligned}
  \tag*{\((\theequation)\)} % <- prints the number inside the box
  \]
\end{tcolorbox}\noindent
where the semi-classical reference value is given by\footnote{Note again the difference of pre-factor, which is again given by $3^{5/12}$ as compared to ~\cite{He:2022wwy} due to the same (but inverse) pre-factor appearing in $M_{\star,\mathrm{SC}}$.}~\cite{He:2022wwy}
\bea
&&T_{\rm max, SC}=\frac{(262440000 \sqrt{2})^{1/3}}{\pi^{17/6}}\frac{m_p\alpha^{19/3}\varepsilon^{5/6}}{g_\star^{4/3}}\,,\nonumber\\
&\approx & 3.2\times 10^9 \mathrm{GeV}\nonumber\\
&&\times\left(\frac{\alpha}{0.1}\right)^{\!19/3}\! \left(\frac{\varepsilon}{4.2\times 10^{-3}}\right)^{\!5/6}\!\!\left(\frac{g_\star}{106.75}\right)^{-4/3}
\eea

\section{Results}\label{sec:MBhotspots}
We will now study specific examples of memory-burden-like transfer functions $\eta_M$ and $\eta_T$ and describe their effect on the hot-spot evolution and maximum temperature.

To do so, we will consider the two extreme cases mentioned above: First, the case where the MB kicks in and leads to constant rates and Hawking temperature, and second, the case of a memory-burden rate that remains self-similar. We stress that although the second case might not be well motivated from a theoretical point of view (see Ref.~\cite{Dvali:2025ktz}), it constitutes a good illustrative example of an evaporation rate that would be significantly suppressed compared to the semi-classical case, while still evolving in time. We emphasise that other cases, such as the smooth transition between the SC and MB regimes~\cite{Dvali:2025ktz, Montefalcone:2025akm}, could easily be studied using the formulas above and lead to substantial variations of the results presented here.
Before continuing, we stress that the hot spot that forms during the semi-classical phase of evaporation is not relevant to the present discussion, though a hot spot would indeed be supported during this period. This is because the semi-classical evaporation rate is large enough to deposit energy rapidly into the surrounding plasma, enabling thermalisation and diffusion. Indeed, during the semi-classical phase, the evaporated energy was sufficiently large to deposit energy quickly into the surrounding plasma due to rapid thermalisation and diffusion thus supporting the formation of a hotspot. 
% Once the MB is at play, the time scales of the problem are stretched 
Once the MB becomes active, the time scales are significantly lengthened. Evaporation takes longer, the flux of particles emitted by the black hole is drastically lowered, and the black hole is unable to maintain any hot spot formed in the semi-classical era. 
% Evaporation takes longer, the flux of particles emitted by the black hole is significantly lowered, and the black hole is not able to maintain any hot spot that was formed initially. 
Therefore, it is sensible to think of a hot spot formed during the memory burden era as a PBH of initial mass $M\approx q M_{\rm ini}$. 
% Therefore, it makes sense to think of a hot spot formed during the memory burden era as formed by a PBH of initial mass $M\approx q M_{\rm ini}$.
\subsection{Rigid Memory-Burden}
We now consider the original MB scenario, in which, after the mass passes the threshold $M<qM_{\rm ini}$, the evaporation rate and Hawking temperature become exactly constant. This corresponds to the choice of transfer functions exhibited in Eqs.~\eqref{eq:etaM} and \eqref{eq:etaT}. 

\begin{figure}
    \centering
\includegraphics[width=\linewidth]{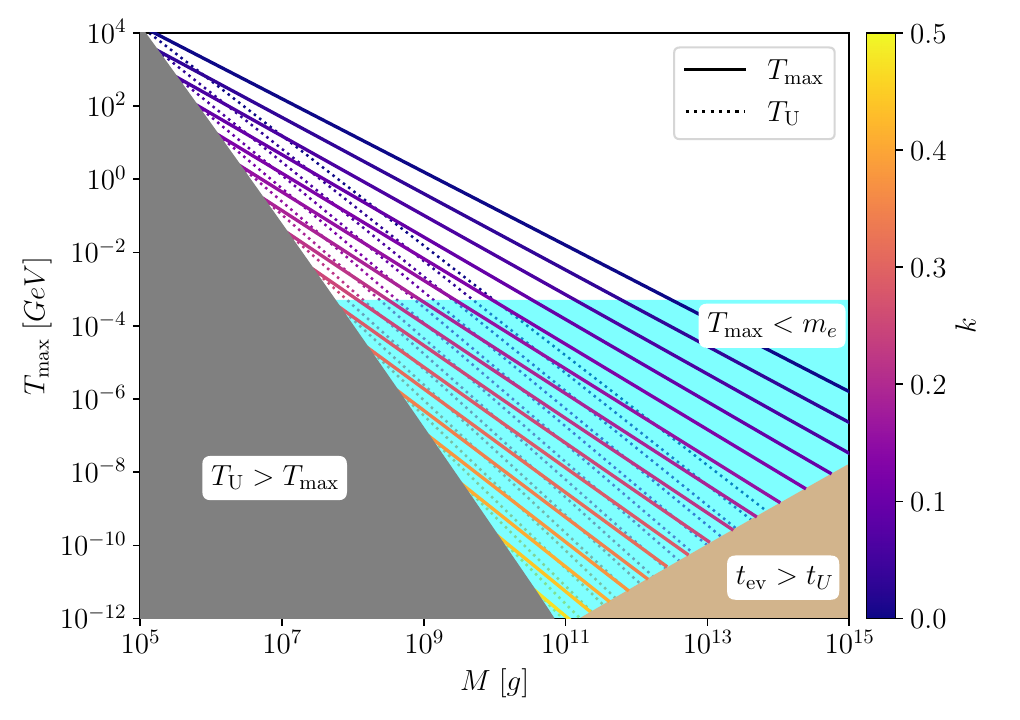}
    \caption{\label{fig:rigid} \footnotesize Hot-spot core temperature (plain coloured lines) compared to the Universe's temperature at the time of evaporation (dotted lines) for different values of $k$, and as a function of $M=qM_{\rm ini}$. Regions where hot spot core is colder than the ambient Universe, and where the PBH is stable on cosmological time scales are shaded in grey and brown, respectively.}
\end{figure}

After the MB kicks in, all the time and energy scales involved in the problem freeze. The shape of the hot spot is thus established while $M\approx qM_{\rm  ini}$ and does not evolve further, until the evaporation is over and diffusion dissipates the hot spot. Due to the large MB suppression of the rate, the core of the hot spot is significantly colder compared to the SC case. In FIG.~\ref{fig:rigid}, we represent the value of this maximum temperature for $q=0.5$ and various values of the parameter $k$. From the figure, one can see that the hot-spot temperature drops rapidly below the electron's mass (cyan-shaded region). This is the threshold at which electrons decouple from the plasma surrounding the black hole. Beyond this threshold it is unclear if any hot spot could form at all since the derivations in this paper depend on established results that assume local thermal equilibrium.
% } Below this point, electrons are decoupled from the plasma surrounding the black hole, and it is not clear that a hot spot can form in the first place at all, as all the results derived in this paper rely on results established assuming a local thermal equilibrium.
Therefore, it seems that the effect of a substantial memory burden is to prevent the formation of a hot spot in that regime. In FIG.~\ref{fig:rigid}, the grey area corresponds to the region of parameter space where the temperature in the ambient universe is greater than the hypothetical hot-spot temperature. Consequently, no temperature gradient can form in this region either.
% On that plot, the grey area denotes the region of parameter space where the temperature in the ambient universe is larger than the hypothetical hot-spot temperature, meaning that no temperature gradient can form in this regime either. 
The tan region also excludes parameters for which the PBH is longer-lived than the Universe. In that case, it thus appears that the vanilla memory-burden scenario does not lead to the formation of a hot spot unless the mass of the black hole is between $10^5\mathrm{g}\lesssim qM_{\rm ini}\lesssim 5\times 10^{12}\mathrm{g}$ and the MB exponent $k\lesssim 0.3$. This excludes, in particular, hot-spot formation in cases with non-zero integer values of $k$.
\begin{figure}
    \centering
\includegraphics[width=\linewidth]{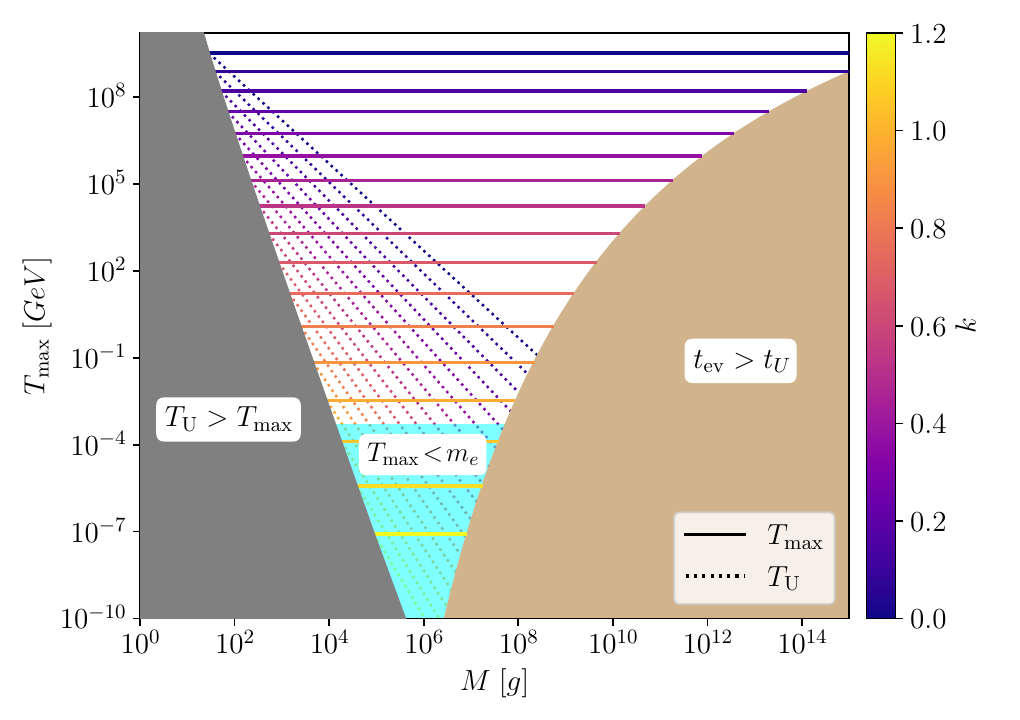}
    \caption{\label{fig:placeholder} \footnotesize Same as in FIG.~\ref{fig:rigid}, but in the case of {\it self-similar} MB rate and Hawking temperature.}
\end{figure}

\subsection{Self-Similar Memory Burden}
If one considers the memory burden suppression to evolve in a self-similar manner with the black hole mass (while the Hawking temperature evolves in an SC manner), one obtains a suppression of the rate that does suppress the hot-spot core temperature, but still allows it to grow over time. This corresponds to setting $qM_{\rm ini}\to M(t)$ in Eqs.~\eqref{eq:etaM} and \eqref{eq:etaT}. In that case, the evolution of the maximum temperature is presented in FIG.~\ref{fig:placeholder}. As expected, as the mass of the PBH decreases, the rate suppression also decreases, and the Hawking temperature is still increasing in a SC manner, leading to a hot spot that can still be much hotter than the ambient universe at the time of evaporation. Notably, the maximum temperature of the hot-spot core is independent of the mass, as in the SC case, as the rate has recovered its self-similarity property. In that case, a hot spot is able to form for masses $50\mathrm{g}\lesssim qM_{\rm ini}\lesssim 5\times 10^{15}\mathrm{g}$ and the MB exponent $k\lesssim 1$.

\section{Summary and Discussion}\label{sec:conclusion}
In this paper, we explored the effect of a modification of the PBH evaporation rate and Hawking temperature evolution on the formation of hot spots, as compared to results established in the semi-classical limit. We introduced general transfer functions $\eta_M$ and $\eta_T$ that encapsulate this modification and derived expressions for the temperature radial profile of the thermal plasma around an evaporating PBH, as a function of these transfer functions and the semi-classical results derived in Ref.~\cite{He:2022wwy}. As expected, we find that due to the memory burden, corresponding to a slowing down of the evaporation process and energy injection, the hot spots predicted are substantially colder than in the semi-classical case. Furthermore, from Eqs.~\eqref{eq:Tcrc} and \eqref{eq:rdec}, it is also easy to see that the overall size of the hot-spot core and outer envelope is also much larger, which is also expected as a colder plasma leads to a smaller energy deposition rate, and therefore a longer mean-free-path for the particles emitted by the BH.
%\lucien{Yes. What about adding: From Eqs.~\ref{eq:Tcrc} and \ref{eq:rdec}, it is also easy to see that the overall size of the hot-spot core and outer envelope is also much larger, which is also expected as a colder plasma leads to a smaller energy deposition rate, and therefore a longer mean-free-path for the particles emitted by the BH.}
For illustrative purposes, we considered the case of a rigid memory burden kicking in at about half the initial mass of the black hole, and showed that a vanilla memory-burden scenario does not lead to the formation of a hot spot unless the mass of the black hole is between $10^5\mathrm{g}\lesssim qM_{\rm ini}\lesssim 5\times 10^{12}\mathrm{g}$ and the MB exponent $k\lesssim 0.3$. This excludes in particular hot-spot formation in cases with non-zero integer values of $k$. For further comparison, we also considered the case where the memory-burden modified evaporation rate would be self-similar, with its expression varying with the PBH mass, while its Hawking temperature would follow the SC evolution.
% We also considered, for comparison, the case where the memory burden rate would be self-similar, with its expression varying with the PBH mass, while its Hawking temperature would follow the SC evolution. 
In that case, despite the large suppression of the PBH evaporation rate, the formation of a hot spot is possible as long as the power-law exponent used in the MB suppression factors is smaller than unity.

As shown in numerous studies~\cite{Gunn:2024xaq, Altomonte:2025hpt, Hamaide:2023ayu, Gross2025}, precise knowledge of the PBH hot-spot evolution is crucial for understanding the imprint of evaporating PBHs in cosmological data. Thanks to the generality of our derivation, we expect that any future derivation of the evaporation rate and Hawking temperature that may be intermediate between the cases we considered in this paper could easily be studied in the future. One such possibility could be the case of a smooth transition between the SC and MB regimes studied in Refs.~\cite{Dvali:2025ktz, Montefalcone:2025akm}. As future research uncovers a more comprehensive understanding of the memory burden and its quantitative effect on the PBH evaporation rate and Hawking emission spectrum, these results will thus allow the community to refine our knowledge of the thermal plasma morphology during the evaporation. 

\section*{Acknowledgements}
The authors would like to thank Clelia Altomonte, Sebastian Zell, Yann Mambrini, Mathieu Gross, and  Jordan Koechler for useful discussions. NL would like to thank Jane Levy and Sayaka Levy for their ongoing support. The work of LH was supported by the STFC under UKRI grant ST/X000753/1. LH acknowledges support by Institut Pascal and the P2I axis of the Graduate School of Physics during the Paris-Saclay Astroparticle Symposium 2025, as well as the CNRS IRP UCMN. The work of NL was funded by the King’s Undergraduate Research Fellowships programme (KURF) at King’s College London.

\vfill

\bibliographystyle{apsrev4-1}
\bibliography{main}

\clearpage

\appendix

\end{document}